\documentclass[runningheads]{llncs}

\usepackage{tikz}
\newcommand*\circled[1]{\tikz[baseline=(char.base)]{
          \node[shape=circle,draw,inner sep=2pt] (char) {#1};}}

\usepackage{hyperref}      
\usepackage{graphicx}
\usepackage{caption}
\usepackage{subcaption}
\usepackage{listings}
\usepackage{xcolor}
\usepackage{algorithmic}
\usepackage{amsmath,amssymb,amsfonts}
\usepackage{textcomp}
\usepackage{float}
 \usepackage{relsize}
\usepackage{url}
\usepackage{hyperref}
\usepackage[htt]{hyphenat}

\definecolor{codegreen}{rgb}{0,0.6,0}
\definecolor{codegray}{rgb}{0.5,0.5,0.5}
\definecolor{codepurple}{rgb}{0.58,0,0.82}
\definecolor{backcolour}{rgb}{1,1,1}
\newcommand{\todo}[1]{}
\renewcommand{\todo}[1]{{\color{red} TODO: {#1}}}
\newcommand\scalemath[2]{\scalebox{#1}{\mbox{\ensuremath{\displaystyle #2}}}}
\lstdefinestyle{mystyle}{
    backgroundcolor=\color{backcolour},   
    commentstyle=\color{codegreen},
    keywordstyle=\color{magenta},
    numberstyle=\tiny\color{codegray},
    stringstyle=\color{codepurple},
    basicstyle=\ttfamily\footnotesize,
    breakatwhitespace=false,         
    breaklines=true,                 
    captionpos=b,                    
    keepspaces=true,                 
    numbers=left,                    
    numbersep=5pt,                  
    showspaces=false,                
    showstringspaces=false,
    showtabs=false,                  
    tabsize=2
}

\begin{document}
\title{Leveraging MLIR for Loop Vectorization and GPU Porting of FFT Libraries}%
\titlerunning{MLIR for Loop Vectorization and GPU Porting of FFT}
%
\author{Yifei He \and
Artur Podobas  \and
Stefano Markidis }
\authorrunning{Y. He et al.}
\institute{KTH Royal Institute of Technology\\
\email{\{yifeihe, podobas, markidis\}@kth.se}}
\maketitle              
\begin{abstract}
\texttt{FFTc} is a Domain-Specific Language (DSL) for designing and generating Fast Fourier Transforms (FFT) libraries. The \texttt{FFTc} uniqueness is that it leverages and extend Multi-Level Intermediate Representation (MLIR) dialects to optimize FFT code generation. In this work, we present \texttt{FFTc} extensions and improvements such as the possibility of using different data layout for complex-value arrays, and sparsification to enable efficient vectorization, and a seamless porting of FFT libraries to GPU systems. We show that, on CPUs, thanks to vectorization, the performance of the \texttt{FFTc}-generated FFT is comparable to performance of \texttt{FFTW}, a state-of-the-art FFT libraries. We also present the initial performance results for \texttt{FFTc} on Nvidia GPUs.
\end{abstract}

\section{Introduction}
Discrete Fourier Transforms (DFT) and their efficient formulations, called Fast Fourier Transforms (FFT), are a critical building block for efficient and high-performance data analysis and scientific computing. In a nutshell, DFTs allow for transforming a digital signal in time, ingested as an input array, into its components in the spectral domain, typically as a complex-value output array. Among several applications, DFTs are widely used for signal processing, e.g., decomposing a signal into its spectral components, or solving Partial Differential Equations (PDE). For instance, the DFT computation is one of the major computational bottleneck in the Particle-Mesh Ewald calculation of the GROMACS, molecular dynamics code~\cite{andersson2022breaking}.

Because of the central role of FFTs in data analysis and scientific computing, several high-performance FFT libraries have been developed. The Fastest Fourier Transform in the West (\texttt{FFTW}) library is among the most used HPC FFT libraries for its performance on serial and parallel systems. In essence, \texttt{FFTW} is a source-to-source compiler emitting a C code to express an FFT library optimized for a given system. However, the FFTW core was first designed and implemented with compiler technologies, nowadays outdated and support only multicore CPUs and not GPUs. On the other hand, the open-source compiler infrastructure is evolving fast. For instance, LLVM became the industry standard for compiler infrastructure, which is the general Intermediate Representation (IR) and supports many hardware back-ends, including for instance GPU programming.  More recently, Multi Level IR (MLIR) introduced the concept of multiple abstraction levels and made it easy to apply high-level domain-specific transformations. All these new efforts in the compiler area can be used to develop a modern FFT libraries that can support heterogeneous computing, multiple hardware backends, including accelerator support. 

\texttt{FFTc} is a new Domain-Specific Language (DSL) built on top of MLIR and LLVM to generate high-performance portable FFT libraries~\cite{he2022fftc}. Differently from \texttt{FFTW}, \texttt{FFTc} can leverage new compiler technologies that allows for a seamless usage of vectorization capabilities and GPU porting. In this work, we present and discuss the new FFTc developments on enabling automatic loop vectorization on CPUs and automatic porting to Nvidia GPUs. To enable these functionalities efficiently, a new data layout for complex-value data and an algorithmic formulation using sparse computation (as opposed to previous FFTc dense computation). The work, presented in this paper, makes the following contributions:
\begin{itemize}
\item We introduce a methodology to convert the complex data type to a first-class type supported by LLVM and hardware ISA, making it possible to apply target-specific optimizations such as vectorization on complex data in MLIR. 
\item We discuss the whole compilation transformation pipeline to enable loop vectorization and GPU usage.
\item We showcase \texttt{FFTc} portability: \texttt{FFTc} can generate efficient code for exploiting vectorization on CPUs and porting FFT libraries to GPU, with a single input source code. 
\end{itemize}

\section{FFTc: An MLIR Dialect for FFT Development}
 Our goal when designing \texttt{FFTc} is to  mask out the hardware details and apply high-level domain-specific optimizations automatically, meanwhile targeting multiple different backends (CPUs/GPUs/etc.) without changes of the source code. 
 
\begin{figure}[h]
    \centering
    \includegraphics[width=0.8\textwidth]{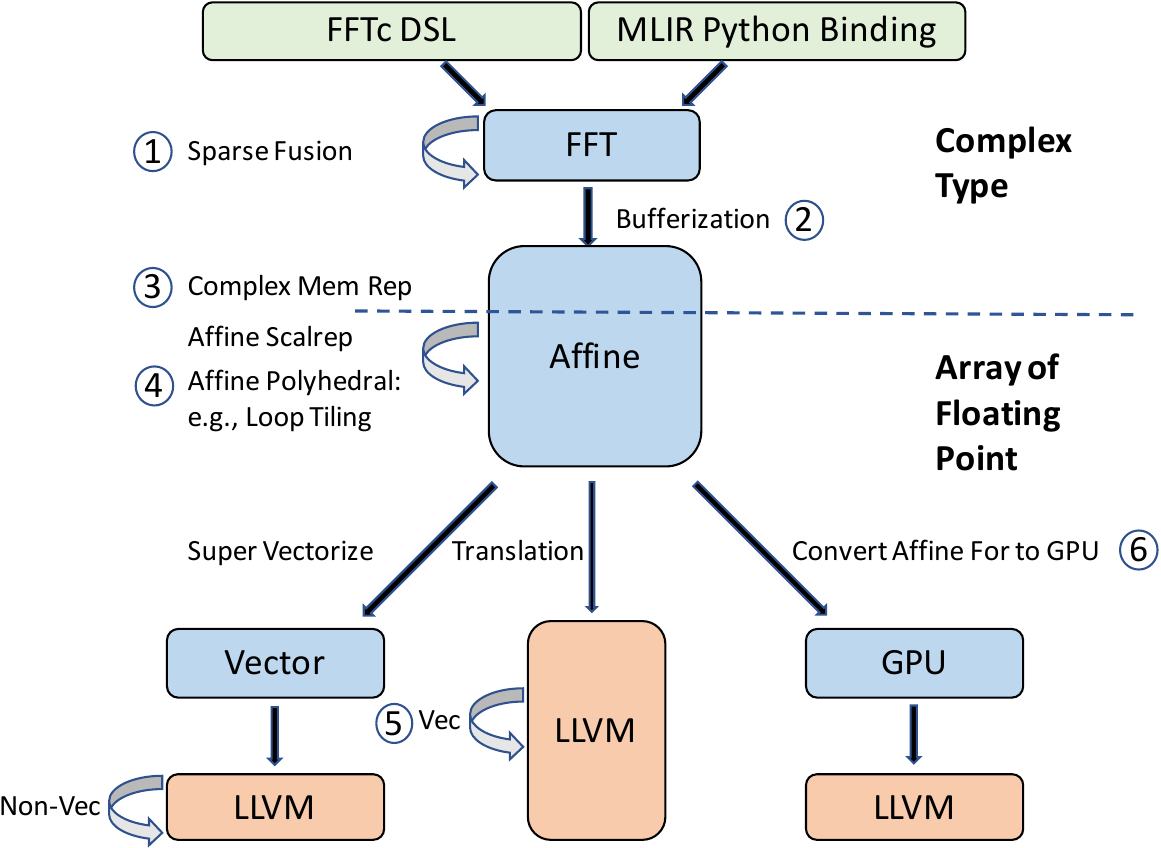}
    \caption{FFTc Compilation Pipeline}
    \label{fig:FFTcwPipeline}
\end{figure}

 We provide an overview of the \texttt{FFTc} compilation pipeline in Fig.~\ref{fig:FFTcwPipeline}. The approach consists a combination of three different components:
\begin{enumerate}
    \item A declarative DSL that operates on tensors (we formulate the FFT algorithm as a factorization of several matrices) and uses tensor products and matrix multiplications. The framework frontend component is shown in the green blocks of Figure~\ref{fig:FFTcwPipeline}. Alternatively, we can also use the MLIR Python binding to generate MLIR directly.  
    \item A MLIR dialect with high-level domain-specific FFT semantics, based on Static Single Assignment (SSA) form. 
    \item A progressive lowering compilation pipeline, which consists of high-level domain-specific optimizations in MLIR and target-specific transformations in MLIR and LLVM. The MLIR dialects and transformations are shown in the blue boxes of Fig.~\ref{fig:FFTcwPipeline}, and the LLVM compilation parts are orange. The detailed description and code examples of FFTc DSL and FFT dialect in MLIR can be found in our prior work~\cite{he2022fftc}.
\end{enumerate}

\section{FFTc Transformations, Loop Vectorization \& GPU Porting}
In this paper, we describe the new FFTc developments in the compilation to enable loop vectorization and GPU porting. In FFTc, we use a tensor formalism expressing FFT algorithms as the factorization of the DFT matrix into sparse matrices. This approach is widely used in developing FFT libraries, such as SPIRAL~\cite{franchetti2018spiral} and Lift~\cite{kopcke2019generating}. For instance, using this formulation, the Cooley-Tukey general-radix decimation-in-time algorithm for an input of size $N$ can be written as: 
\begin{equation} \label{eq:Recursive FFT}
\scalemath{0.85}{
\operatorname{DFT_N} = (\operatorname{DFT_K}\otimes \operatorname{I_M})\operatorname{D^N_M}(\operatorname{I_K} \otimes \operatorname{DFT_M})\operatorname{\Pi_K^N} \quad \text{with} \quad N = MK,
}
\end{equation} 
where $\operatorname{\Pi_K^N}$ is a stride permute operator and $\operatorname{D_M^N}$ is a diagonal matrix of \textit{twiddle} factors.
We present now the new developments following the compilation pipeline, presented in Fig.~\ref{fig:FFTcwPipeline} as differed numbered phases. \\

\noindent \textbf{\circled{1} FFTc Sparse Fusion Transformation.} The first implementation we demonstrated in the DSL used dense matrix representation and computation~\cite{he2022fftc}, e.g., we perform calculations also for zero matrix entries. In this new FFTc version, we perform the computation in sparse format to achieve the $\mathcal{O} (N \log N)$ complexity of FFT. The FFT dialect closely represents the semantics of the mathematical formula, carrying high-level information about the FFT computation. Also, the FFT dialect works on tensor values that are immutable and without side effects, which brings convenience for compiler analysis and transformations. Therefore, we perform the Sparse Fusion Transform (SFT) on FFT dialect. SFT uses the pattern match and rewriting mechanism in MLIR to fuse several FFT operators into one. As shown in Table \ref{table:mapDSL2Sparse}, the FFT computation pattern Y = ($A_m$ {$\otimes$} ${I_n}$) {$\cdot$} X is fused into one operator \texttt{FusedMKIV}. Here $M$ stands for matrix, $K$ for kronecker product, $I$ is identity matrix, and $V$ means vector.  \\

\begin{table}[t]
\centering
\caption{Sparse Fusion and Bufferization Transform.}
\begin{tabular}{c c l}
\textbf{FFTc DSL Pattern}  & \textbf{Sparse Fusion}
& \textbf{\ \ \ \ Bufferization}  \\
        \hline
Y = ($A_m$ {$\otimes$} ${I_n}$) {$\cdot$} X 
& FusedMKIV(A, n, X) 
&  for($i=0; \ i<n; \ i++$)\\
& & \ \ Y[$i : n : i+m*n-n$] = \\
& & \ \  A$*$(X[$i : n : i+m*n-n$])\\

Y = (${I_m}$ {$\otimes$} $A_n$) {$\cdot$} X 
& FusedIKMV(A, n, X) 
& for($i=0; \ i<m; \ i++$) \\
& & \ \ Y[$i*n : 1 :i*n+n-1$] =\\
& & \ \  A$*$(X[$i*n : 1 : i*n+n-1$])
\\
({$\Pi_m^{mn}$} {$\otimes$} {$I_k$}) {$\cdot$} X & 
FusedPKIV(m, mn, k, X) &
for($i=0; \ i<m; \ i++$) \\
& & \ \  for($j=0; \ j<n; \ j++$) \\
& & \ \ \ \  Y[$k*(i+m*j) : 1 : k*(i+m*j)$] = \\
& & \ \ \ \ X[$k*(n*i+j) : 1 : k*(n*i+j)$]
\\

{$D^n_m$} {$\cdot$} X  &
Mul(TwiddleCoe, X) &
 for($i=0; \ i<m; \ i++$) \\
&& \ \ Y[$i$] = $D^n_m$[$i$] * X[$i$]) \\

{$\Pi_m^{mn}$} {$\cdot$} X  & Permute(m, mn, X) 
&  for($i=0; \ i<m; \ i++$) \\
& &  \ \  for($j=0; \ j<n; \ j++$) \\
& & \ \ \ \  Y[$i+m*j : 1 : i+m*j$] = \\ & & \ \ \ \ A$*$(X[$n*i+j : 1 : n*i+j$])
 
 \\ [2ex]
\end{tabular}
\label{table:mapDSL2Sparse}
\begin{tabular}{c c}
\end{tabular}
\end{table}

\noindent \textbf{\circled{2} Bufferization: Lower to Affine and MemRef Dialect.} After the high level transformations on FFT dialect, we apply bufferization, the FFT dialect operations with tensor semantics are lowered down to explicit loops with MemRef semantics. In MLIR, the Memref dialect provides us with a method to manipulate the allocation and data layout of the memory pointed by the memref type (a specialized data type used to represent multi-dimensional arrays or buffers with memory layout information).

 The MemRef dialect represents the lower-level buffer access and builds a bridge to the actual computer memory.  The lowering matches patterns of individual FFT dialect operations, rewriting them with explicit affine loop nests to implement the computations. The scalarized tensor arithmetic operations are performed by corresponding operations in the Complex dialect. The pseudo code of the FFT operations after bufferization, are demonstrate in Table~\ref{table:mapDSL2Sparse}. We can see that the computation is already sparsified. We got some inspiration from SPIRAL~\cite{franchetti2018spiral} for the sparse fusion and bufferization work.\\

\noindent \textbf{\circled{3} Conversion of Complex Data to an Array of Floating-point.} The FFT algorithms operate on complex numbers and it is critical to have high-performance data access to complex-value arrays. The complex dialect in MLIR is used to hold complex numbers and perform complex arithmetic operations. The complex data type is more explicit when representing computation workloads, also more convenient for domain-specific transformations since the aggregated complex data is wrapped as a single unit. So we apply all the domain-specific transformations to the complex data type. However, the complex data type is neither a first-class type in LLVM nor widely supported by hardware instructions. Also, in MLIR, some dialects cannot work with the complex data type, e.g., Vector dialect. The Vector dialect is a low-level but still machine-agnostic dialect for virtual vector operations~\cite{MLIRVectorCodeGen}. The virtual vector operations will map closely to LLVM IR and, eventually, hardware vector instructions. 

To perform vectorization in MLIR, we convert the complex type to an array of floating point data types. We introduce a conversion pass \texttt{fft-convert-complex-to-floating} and a rewriting pass \texttt{{fft-complex-mem-rep}}. These two passes apply conversion patterns to convert the operations of complex dialect and other dialects’ operations on complex data to memory access operations on an array of floating point data. For instance, as shown in the list \ref{lst:Complex2Floating}, after conversion the complex data type is eliminated.

\begin{lstlisting}[language=Python, float=*, extendedchars=true, captionpos=b, caption= Converting Complex Dialect Operations, label={lst:Complex2Floating}, literate={·}{{$\cdot$}}1 {⊗}{{$\otimes$}}1]
From: %12 = "complex.create" %10, %11 : complex<f64>

To:   %13 = "memref.alloc"() : memref<2xf64>
      "affine.store" %14, %13[0] : memref<2xf64>
      "affine.store" %15, %13[1] : memref<2xf64>
      %16 = builtin.unrealized_conversion_cast %13 : memref<2xf64> to complex<f64>
\end{lstlisting}
We can also change the data layout of the complex array here by setting a flag to the pass: currently, we can switch between the \textit{Interleaved} and \textit{Split} modes. In the \textit{Interleaved} mode, the real and imaginary parts of a complex number are located in consecutive memory locations~\cite{popovici2017mixed} ~\cite{andersson2023case}. On the other hand, the \textit{Split} data format stores the real and imaginary components as two disjoint sequences. \\

\noindent \textbf{\circled{4} Polyhedral Transformations in the Affine Dialect.} The Affine dialect is a simplified polyhedral representation designed to enable progressive lowering~\cite{lattner2021mlir}. We utilize the transformations in the Affine dialect to explore the loop optimization opportunities, such as fusion, tiling, and vectorization. The automatic loop fusion at this stage cannot generate the most optimal fused loops, so we rely on the Sparse Fusion pass, which works on the high-level tensor data. 

The loop optimization we develop in FFTc are:
\begin{itemize}
\item \textbf{Loop Tiling.} We need to set the loop tiling size as a hyperparameter to the \texttt{{affine-loop-tile}} pass. There are two ways of setting the tiling size: exact tile size or indicating the target cache volume. We found the latter ones performed better for our FFT code generation.
\item \textbf{Loop Vectorization.} Affine dialect's \texttt{{affine-super-vectorize}} pass is designed to generate virtual vector operations out of loops. At this stage, the FFT transforms mentioned above have already generated parallel loops without dependencies which may prevent vectorization. The \texttt{{affine-super-vectorize}} needs to determine the most optimal loop dimension and virtual vector size to perform vectorization, either generated automatically by heuristics or set as hyperparameters by the developer.

After the naive vectorization, we run the \texttt{{test-vector-transfer-lowering-patterns}} pass to apply optimization patterns on the vector operations. 

\end{itemize}
\begin{figure}
     \centering
     \begin{subfigure}[b]{0.32\textwidth}
         \centering
         \begin{lstlisting}[ language=Python, extendedchars=true, captionpos=b]
for(i=0;i<M;i++)
  for(j=0;j<N;j++)
    c[i][j]=
    a[i][j]+
    b[i][j];\end{lstlisting}
         \caption{Scalar Loop}
         \label{fig:Scalar Loop} 
     \end{subfigure}
     \hfill
     \begin{subfigure}[b]{0.32\textwidth}
         \centering
         \begin{lstlisting}[ language=Python, extendedchars=true, captionpos=b]
for(i=0;i<M;i++)
  for(j=0;j<N;j+=4)
    c[i][j:j+3]=
    a[i][j:j+3]+
    b[i][j:j+3];\end{lstlisting}
         \caption{Inner Loop Vectorized}
         \label{fig:Inner Loop} 
     \end{subfigure}
     \hfill
     \begin{subfigure}[b]{0.32\textwidth}
         \centering
         \begin{lstlisting}[ language=Python, extendedchars=true, captionpos=b]
for(i=0;i<M;i+=4)
  for(j=0;j<N;j++)
    c[i:i+3][j]=
    a[i:i+3][j]+
    b[i:i+3][j];\end{lstlisting}
         \caption{Outer Loop Vectorized}
         \label{fig:Outer Loop} 
     \end{subfigure}
     \hfill
        \caption{Loop Vectorization Example ( Vector Length = 4 )}
        \label{fig:Loop Vectorization}
\end{figure}

\subsection{LLVM Loop Vectorizer}
After the FFT code is lowered to LLVM IR, we can leverage the LLVM pipeline to further optimization. We specifically explore opportunities in vectorization, one of the most critical optimizations for performance. As shown in Fig.~\ref{fig:FFTcwPipeline}, for this case we bypass the MLIR vectorizer. 

There are two different vectorizers in the LLVM pipeline: SLP~\cite{rosen2007loop} and VPlan~\cite{tian2017llvm} vectorizers. From a user's perspective, SLP is a innermost loop vectorizer, as shown in Fig. \ref{fig:Inner Loop}, and VPLAN vectorizes the outermost loop, demonstrated in \ref{fig:Outer Loop}.

The current heuristic in LLVM loop vectorizer to choose the vector length will block AVX512 instructions for Intel CPUs, due to the probable frequency drop, which may negate the performance gains, especially for non-computation-intensive workloads. In our FFT computation, wider vectors brings more performance, so we modified the x86 target configuration and \texttt{PreferVectorWidth} heuristic in LLVM to enable AVX512 code generation.

\begin{figure}
     \centering
     \begin{subfigure}[b]{0.45\textwidth}
         \centering
         \includegraphics[width=\textwidth]{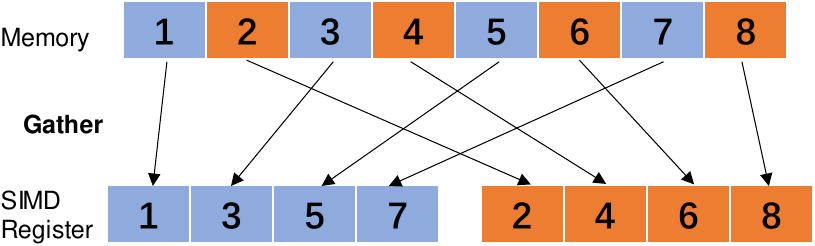}
         \caption{Directly Load Complex Data Using Gather Instructions}
         \label{fig:Gather} 
     \end{subfigure}
     \hfill
     \begin{subfigure}[b]{0.45\textwidth}
         \centering
         \includegraphics[width=\textwidth]{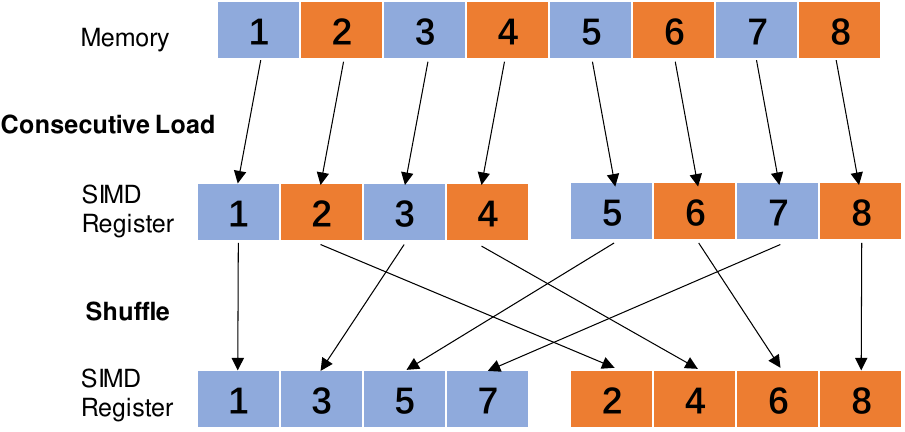}
\caption{Optimized Interleaved Memory Access}
\label{fig:Interleave}
     \end{subfigure}

        \caption{Pack Complex Data into SIMD Registers during Auto-vectorization}
        \label{fig:Memory Access}
\end{figure}

\noindent \textbf{\circled{5.1} Interleaved Memory Access and Innermost Loop Vectorizer: SLP.} The complex numbers are stored under interleaved data layout in most cases, such as the complex data type in C++: \texttt{std::complex}. Interleaved means that a complex number's real and imaginary parts are stored consecutively in memory. However, this raises a challenge for effective memory access using SIMD instructions. For example, strided load is needed to pack the real/imaginary parts of multiple complex numbers into a SIMD register, which may end up using expensive gather instructions. An example is shown in Fig.~\ref{fig:Gather}.

An interleaved memory access optimization is available in LLVM's loop vectorizer, and we enable this by explicitly setting the corresponding flag to the loop vectorize pass. After the optimization, two consecutive SIMD loads will be generated, followed by shuffle operations on these two SIMD registers to replace the gather/scatter, as shown in the Fig.~\ref{fig:Interleave}. The interleaved memory access optimization can work together with the SLP inner loop vectorizer.

SLP (Superword-Level Parallelism) is the default loop vectorizer in the LLVM optimization pipeline, targeting to combine similar independent instructions in the innermost loop into vector instructions. 

\noindent \textbf{\circled{5.2} Outermost Loop Vectorization: VPLAN.} Outermost vectorization can be beneficial for some cases, e.g., the number of iterations in the innermost loop is small. VPlan is a recently introduced LLVM vectorizer. 
A current development effort is ongoing to migrate the loop vectorizer to the VPlan infrastructure and support outer loop vectorization in the LLVM loop vectorizer. Currently, VPlan is a temporary vectorization path. For this reason. we need to set \texttt{enable-vplan-native-path} option to enable it. Also, VPlan only vectorizes the outermost loop with explicit vectorization annotation, e.g., \texttt{\#pragma omp simd}. We modified the execution logic to make it run on our FFT loops.

\subsection{FFTc GPU Code Generation}
As shown in Figure~\ref{fig:FFTcwPipeline}, the progressive lowering above the Affine dialect is hardware target agnostic. Below there are divergent branches to support different hardware targets and heterogeneous computing. To generate GPU code, we lower down to MLIR GPU dialect~\cite{vasilache2022composable}, the retargetable GPU programming model in MLIR. The MLIR GPU dialect can further lowering down to different hardware targets, such as NVIDIA and AMD GPUs. We demonstrate the subsequent GPU code generation using the NVIDIA compilation pipeline. \\

\noindent \textbf{\circled{6.1} Preparation: Partial Lowering Affine Memory Operations.} We generate the GPU kernel from Affine loops using the \textbf{\textit{convert-affine-for-to-gpu}} pass. However, the current implementation of the pass does not support memory access using Affine load/store. For this reason, we introduce a partial lowering pass to convert the Affine memory access to corresponding ones in the Memref dialect, e.g., \texttt{memref.load/store}. Also, we need to register all the Memref used in GPU kernel using \texttt{gpu.host\_register}, to access it from the device. \\

\noindent \textbf{\circled{6.2} Convert Affine Loops to GPU Kernel.} The \textbf{\textit{convert-affine-for-to-gpu}} pass converts each Affine loop nest into a GPU kernel. This pass collects the loop nest's ranges, bounds, steps, and induction variables, then uses them to calculate the grid and block sizes for the GPU kernel. \\

\noindent \textbf{\circled{6.3} CUDA Binary Code Generation.} In the generated code with GPU dialect, the CPU (host) and GPU (device) codes are embedded in a single IR. However, the GPU kernel code is wrapped into specific functions to separately run compilation passes. The GPU device code is lowered to platform-specific dialects, such as the Nvidia NVVM, a compiler IR for CUDA kernels based on LLVM IR. As the last step, the CUDA binary code is generated. Returning to the host side, we lower the host side GPU code to LLVM, then go through the LLVM code generation pipeline. We support both Just-in-Time (JIT) and Ahead-of-Time (AoT) compilation modes for the GPU code generation.

\section{Experimental Setup \& Evaluation}
We evaluate the CPU performance of the FFTc-generated FFT on the Tetralith supercomputer, located at the National Supercomputer Centre in Linköping, Sweden. The Tetralith computing nodes have a dual-socket Intel Xeon Gold 6130 CPU, 96 GB of RAM. 
For GPU tests, we use the Alvis supercomputer in Chalmers, Sweden. Each Alvis' computing node has 4xA40 NVIDIA GPUs, the CUDA version is 12.0.0. The LLVM we use to embed the FFTc was forked from the LLVM main branch on 2022/08/11.  


We run the FFT kernels 1,000 times and calculate the average execution time. We develop a Python script to generate the implementation of the FFT algorithm using our FFTc DSL. The other option is using MLIR Python binding to generate MLIR directly. Albeit our script can generate different FFT algorithm implementations, in this paper, we mainly present the results of the Stockham algorithm.

We evaluate our compiler on double-precision complex-to-complex FFT. Currently, we do not support mixed radix algorithms and the runtime decomposition of DFT as FFTW does. For this reason, in this study, the FFT sizes are all powers of two. We precompile our FFT kernels into a static library and then call it in a C++ file using a similar API as FFTW. The input/output data type for MLIR library code is the member data type, which describes a structured multi-index pointer into memory. To convert it to an externally-facing function and call in C/C++, we use \texttt{llvm.emit\_c\_interface} function to generate wrapper functions that convert a Memref augment to a pointer-to-struct argument.
A wrapper on C++ array pointer is also provided to generate the pointer-to-struct structure equivalent to MLIR Memref when lowering down to LLVM IR.

\begin{figure}[t]
     \centering
     \begin{subfigure}[b]{0.45\textwidth}
         \centering
         \includegraphics[width=\textwidth]{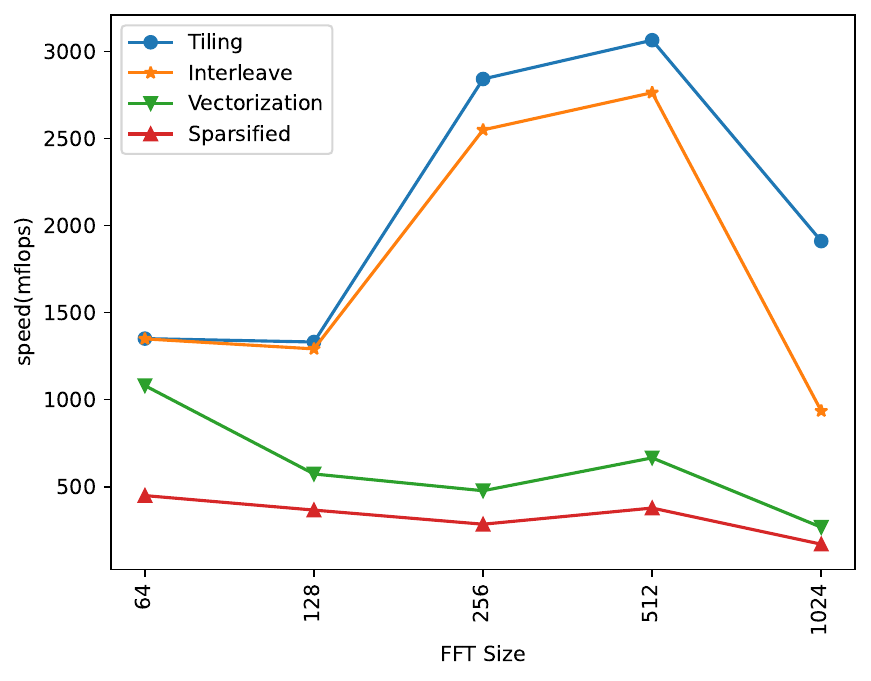}
         \caption{Different and combined optimizations on CPU.}
         \label{fig:PerfOpt} 
     \end{subfigure}
     \hfill
     \begin{subfigure}[b]{0.45\textwidth}
         \centering
         \includegraphics[width=\textwidth]{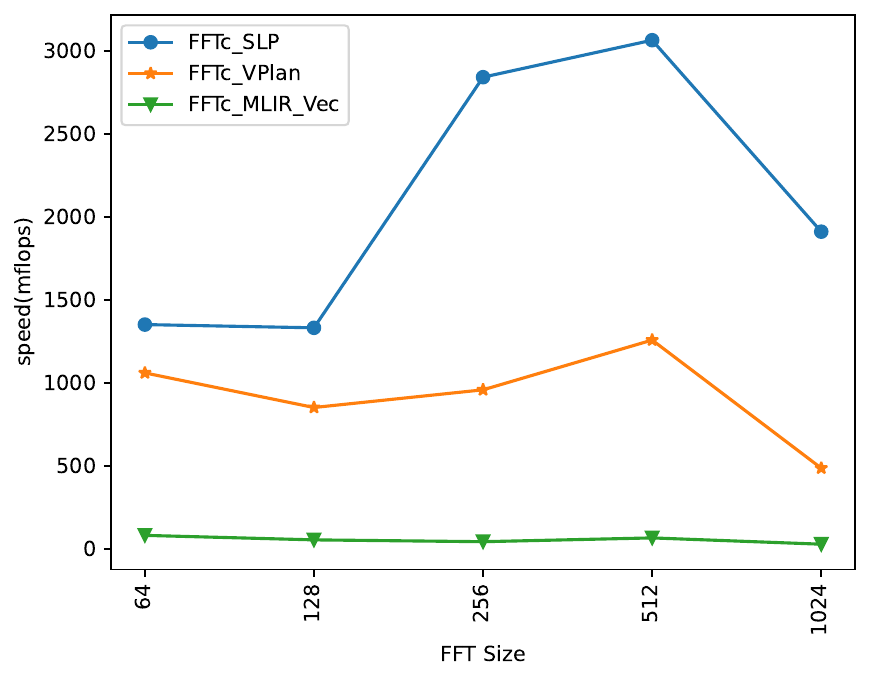}
\caption{Different vectorizers approaches on CPU.}
\label{fig:PerfVec}
     \end{subfigure}
     \hfill
     \begin{subfigure}[b]{0.45\textwidth}
         \centering
         \includegraphics[width=\textwidth]{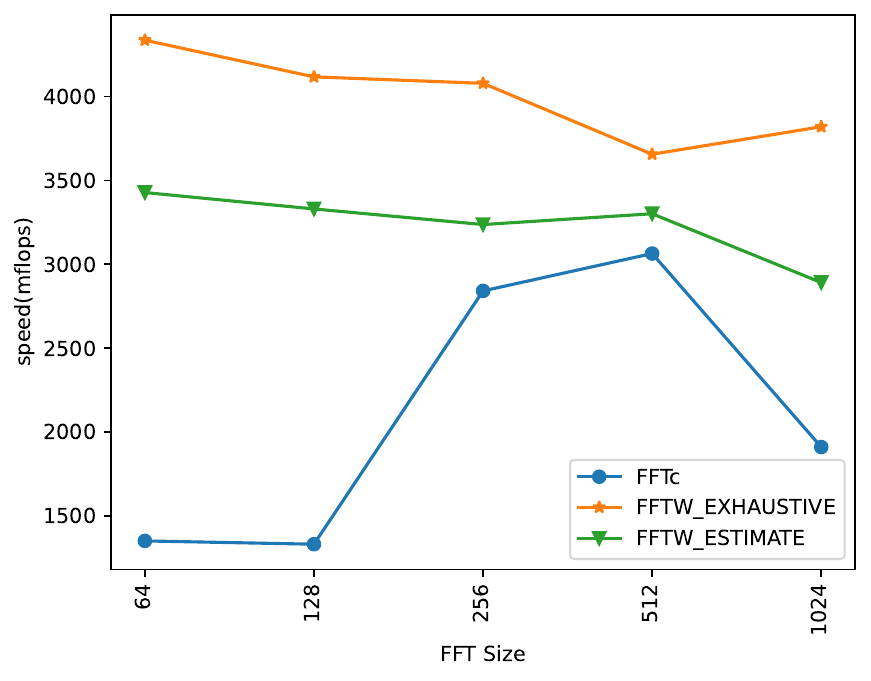}
         \caption{FFTc compared with FFTW on the CPU.}    \label{fig:PerfStateofArt}
     \end{subfigure}
     \hfill
     \begin{subfigure}[b]{0.45\textwidth}
         \centering
         \includegraphics[width=\textwidth]{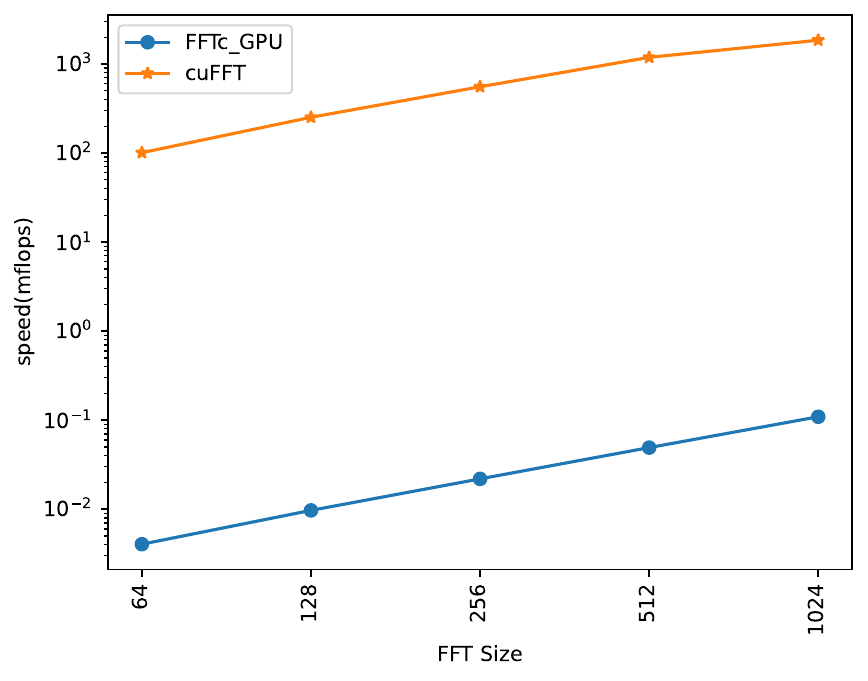}
         \caption{FFTc (naive) compared with cuFFT on GPU.}
        \label{fig:PerfGPU}
     \end{subfigure}
        \caption{Performance Results of FFTc. CPU performance is tested on on Tetralith compute node with Intel Xeon Gold 6130 CPU. GPU performance on Alvis compute node with NVIDA A40 GPU.}
        \label{fig:Performance}
\end{figure}

\noindent \textbf{Results.} As the first step of our work, we verify the correctness of FFTc implementation. The test takes random input vectors as input, with different FFT sizes: the input sizes are the powers of two, from 16 to 4,098. We call the FFTc library from the C++ test case and compare the results with FFTW. The error is calculated as $\frac{|result_{DSL} - result_{Numpy}|}{FFT_{size}}$. In all our runs, the error is smaller than 1e-7. 

 In Fig.~\ref{fig:PerfOpt}, we report performance attained by different optimizations. We run it on FFT sizes of power of two and avoid large FFT sizes since we do not support a runtime planner like FFTW to decompose the FFT into smaller problems, nor support random-sized FFT. We apply the same optimization pipeline and hyperparameters to all different FFT sizes. Performance is demonstrated in MFLOPs/s. The MFLOPs are calculated as $5(N\log_2(N))$, where N is the number of input data points.
 
 The compiler transforms with low or detrimental performance are not reported, and we report only the ones with performance gains. In Fig.~\ref{fig:PerfOpt}, the red bottom line stands for the FFT code after sparse fusion. This optimization leads to up to 1,000x performance improvement compared with the dense computation before sparsify and significantly reduces the compilation time~\cite{he2022fftc}. The dense computation is implemented through dense matrix multiplication, which requires $\mathcal{O}(N^2)$ operations: in this case, the MFLOPS/s needed for specific FFT sizes are different, so we do not compare them with the sparsified ones.
 
 The green line represents the sparsified code with SLP vectorization; it achieves approximately 2x speed-ups for most cases. The most significant performance gain comes from the interleaved memory access optimizations: up to 5x speed up is achieved. Finally, we can gain additional performance with loop tiling, especially for large FFT sizes.

 In Fig.~\ref{fig:PerfVec}, we present the performance with different vectorizers. The green bottom line stands for the MLIR vectorizer. Although it successfully vectorized the code (on virtual vector abstraction), the subsequent lowering passes failed to generate optimal target-specific vector instructions, e.g., in our cases, the memory access instruction \texttt{vector.transfer\_read/write} is scalarized when lowering down. In the current implementation of the MLIR pipeline, only some memory access patterns are efficiently mapped to vector instructions. Further customization and fine-tuning of the vectorizer are needed to generate high-performance code. The other path is to lower the scalar MLIR code to LLVM and utilize LLVM's vectorizers. We can see that the innermost loop vectorizer SLP with interleaved memory access optimization outperforms the outermost loop Vectorizer VPlan. VPlan cannot work with interleaved memory access optimization and generates gather/scatter instructions.

 In Fig.~\ref{fig:PerfStateofArt}, we compare our implementation with the state-of-art library FFTW. We test FFTW with different planner flags, which control the planning process and, therefore, the overall FFTW performance. We choose \texttt{FFTW\_ESTIMATE} and \texttt{FFTW\_EXHAUSTIVE} out of all the options. To pick an optimal plan, \texttt{FFTW\_ESTIMATE} uses a simple heuristic, which requires the least compilation time. While \texttt{FFTW\_EXHAUSTIVE} performs an exhaustive search, it is the most time-consuming option: it will compute several FFTs and measure the execution time and select the algorithm with the best performance.
 
 There is a large performance gap for certain sizes, such as 64 and 128. Part of the reason is that currently, we cannot support mixed radix FFT algorithms, so we cannot decompose the random FFT size into radix sizes that match the memory hierarchy capacities. For instance, size-128 FFT is calculated using radix-2 kernels, which is not optimal for memory access and vectorization. The other reason would be that we did not apply aggressive unroll on small-size kernels. The performance difference between FFTc and FFTW is relatively small for the FFT sizes, which we can decompose into vectorization and memory access-friendly radix sizes. An example is size-256 FFT, which can be split into radix-16 kernels.

Fig.~\ref{fig:PerfGPU} demonstrates the GPU performance result compared with the Nvidia FFT library. The primary purpose is to show FFTc's portability, that we can support multiple hardware targets with a single source code. We have not investigated performance optimization on GPU: currently, the performance difference between cuFFT and FFTc is considerable. In FFTc, each affine loop nest of FFT code is mapped into a GPU kernel: at this point, the Affine loop nest is not optimized to map efficiently to the hierarchical hardware parallelism (grid/block) of a GPU. We plan to do this in the near future, together with other optimizations, e.g., vectorization and memory promotion, to utilize the hierarchical (shared/private) memory in GPU.

\section{Related Work}
The most widely used open-source FFT library, FFTW~\cite{frigo1998fftw}, is essentially an FFT compiler. FFTW is written in Objective Caml to generate Directed Acyclic Graphs (DAG) of FFT algorithms and perform algebraic optimization. FFTW uses a planner at runtime to recursively decompose the DFT problem into sub-problems. This sub-problems are solved directly by optimized, straight-line code generated by a special-purpose compiler called \texttt{{genfft}}~\cite{frigo2004fast}. Another successful library using compiler technology is SPIRAL~\cite{franchetti2018spiral}, which uses a mathematical framework for representing and deriving numerical and scientific algorithms, including FFTs. SPIRAL applies pattern match and rewriting to generate optimal FFT formulation for different hardware, such as SIMD and  multicore systems. Then, SPIRAL maps the matrix formula to high-performance C code. A similar approach to our methodology is used in the Lift framework~\cite{kopcke2019generating} that uses compiler technologies and a mathematical formulation to generate FFT libraries for different hardware, including accelerators. Differently from all these approaches, in our work we build our framework on the top of the recent LLVM and MLIR technologies.

\section{Discussion \& Conclusion}
\texttt{FFTc} is a DSL built upon a series of existing and newly introduced abstractions in MLIR and LLVM. In particular, \texttt{FFTc} allows for decoupling the high-level domain-specific FFT abstractions (tensor operations) and lower-level target-specific abstractions (vector instructions and CUDA primitives). 

In this paper, we described the new \texttt{FFTc} features, including automatic support for loop vectorization and the possibility of generating code for execution on GPUs. In order to enable loop vectorization and GPU porting a number of LLVM and MLIR transformation are needed.

While we show that the \texttt{FFTc} performance on CPU is on-par with the performance of \texttt{FFTW}, the performance of the FFTc-generated code is largely suboptimal to the performance of \texttt{cuFFT}. As a priority for future work, we will investigate the MLIR's vectorizer to generate optimized vector code both for GPU and CPU and improve the Affine loops to be efficiently mapped to CPU's Multicore/SIMD and GPU's workitem/group (thread/block) hierarchical parallelism architecture. In addition, similarly to \texttt{FFTW}, we will investigate the development of a compiler runtime to generate and select optimal candidates of FFT formula decomposition plans.

%
%
%
%
\bibliographystyle{splncs04}
\raggedright
\bibliography{ref}
\end{document}